\documentclass[preprint]{elsarticle}
\usepackage{amsmath}
\usepackage{siunitx}
\usepackage{graphicx}
\usepackage[font=small,labelfont=bf]{caption}
\usepackage{hanging}

\biboptions{sort&compress}

\begin{document}

\begin{frontmatter}

\title{Analysis and optimization of a multicascade  method for the size fractionation of poly-dispersed particle systems via sedimentation or centrifugation}

\author[1]{Heng Li}
\author[1]{Lucas Beetsma}
\author[1]{Suriya Prakash}
\author[2]{Maurice Mikkers}
\author[1]{Lorenzo Botto\corref{cor1}}
\ead{l.botto@tudelft.nl}
\affiliation[1]{organization={Process and Energy Department, 3ME Faculty of Mechanical, Maritime and Materials Engineering, TU Delft},
            postcode={2628 CD Delft}, 
            country={The Netherlands}}
\cortext[cor1]{Corresponding author}
\affiliation[2]{organization={Maurice Mikkers B.V.},
            addressline={Newtonstraat 296}, 
            postcode={2562 KZ Den Haag}, 
            country={The Netherlands}}

\begin{abstract}
Sedimentation and centrifugation can be used to sort particles by size, using a multistep (multicascade) method in which particles in the sediment are removed and the content of the supernatant is processed again, repeating the cycle several times.  This paper proposes a theoretical analysis of this process, based on a one-dimensional model, with a view to identify parameters that are optimal to obtain a relatively monodispersed suspension, starting from a log-normal particle size distribution. We found that a rational choice of the sedimentation/centrifugation  time enables to control the amount of particles outside of the desired size range (impurities). Surprisingly, the multistep method does not converge, as multiple steps are worse than 2 steps. Band sedimentation, in which a particle-rich layer is overlaid on clear fluid, offers substantial benefits in terms of impurity reduction with respect to starting from a completely mixed situation. An application to graphene fractionation is discussed.
      
\end{abstract}


\end{frontmatter}

\section{Introduction}
\label{introduction}
Dispersed liquid-solid and liquid-liquid systems are liquids containing suspended solid or liquid particles. Examples of such systems are cell broths, protein solutions, nanomaterial suspensions and emulsions \citep{mewis2012colloidal,lee2020comprehensive}. A very common technique to separate the dispersed and continuum phases in these systems is the application of a gravitational or centrifugal field. If the densities of the dispersed and continuous phases are different, the applied field will drive the translation of the dispersed phase with respect to the continuum phase, so that the dispersed phase can be removed from the fluid \citep{abeynaike2012experimental,majekodunmi2015review}. Decanters, centrifuges, ultra-centrifuges and gravity settlers, are examples of separation units that work based on this method.

In addition to separate particulate materials from the liquid, sedimentation and centrifugation can also be used to fractionate particles in size classes, or to make the particle size distribution more monodisperse.  This is a crucial outcome, as the performance of colloidal materials and nanomaterials is extremely dependent on particle size and obtaining particles with controlled polydispersity has practical and economical advantages.  For example, the cost of many nanomaterials depends markedly on how monodispersed the sample is, and in fact for applications demanding high purity fractionation is  necessarily applied to samples provided by suppliers. There are of course many methods for particle fractionation, for example field-flow fractionation \citep{schimpf2000field}, hydrodynamic chromatography \citep{striegel2012hydrodynamic} and filtration \citep{chen2015size}, but centrifugation and sedimentation are unique in their ability to handle large or coarse samples and are quick and easy to apply. 

\begin{figure}
    \centering
    \includegraphics[width=0.7\textwidth]{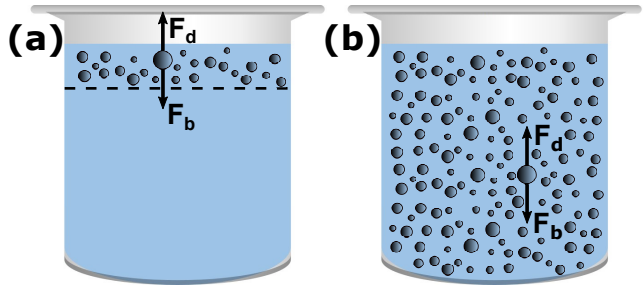}
    \caption{Sketches of settling particles under the balance of body force ($\mathbf{F_b}$) and Stokes drag ($\mathbf{F_d}$) in (a) band sedimentation and (b) homogeneous sedimentation.}
    \label{fig1}
\end{figure}

\begin{figure}
    \centering
        \includegraphics[width=1.0\textwidth]{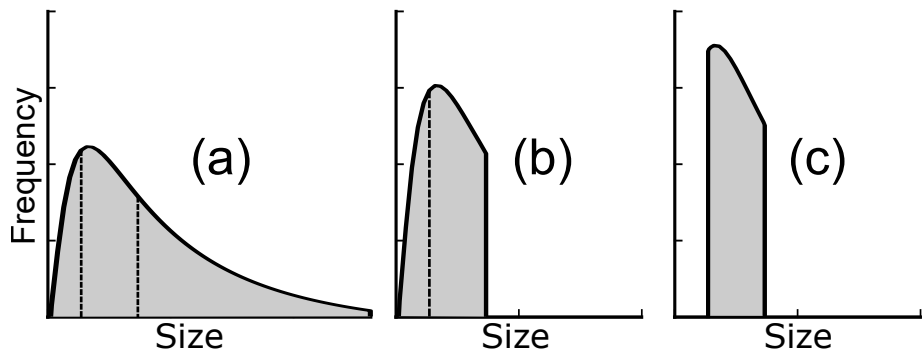}
        \caption{Ideal case of the isolation of certain size range by successive steps. Frequency distributions (a) initially, (b) after first step and (c) after second step.}
        \label{fig2}
\end{figure}

The method through which size fractionation can be obtained by sedimentation or centrifugation is a multistep method called, in the context of centrifugation, differential  or multicascade centrifugation. The method consists in removing the sedimented layer from the supernatant after a specified centrifugation time, then the supernatant is centrifuged again. The process is then repeated several times until it is expected, based on some criteria, that the fractionated sample contains only particles of a given size interval, and no others. The decision for the centrifugation parameters, for instance centrifugation time and speed, is typically based on trial and error, or following literature protocols that are often themselves based on empirical observations. The question that this paper would like to address is whether this choice can be made rationally based on a mathematical model that provides the particle size distributions in the supernatant and sedimented layer at any given time. This mathematical model will furnish also the amount of impurities. When deciding for example the centrifugation times to isolate a certain interval of the initial size distribution, one indeed must account for the fact that there is a link between the centrifugation time and the amount of impurities obtained. For example, for any choice of centrifugation time, some fine particles will sediment at the bottom of the centrifugation vial together with the coarser particles, so the pellet at the bottom will be a mixture of mainly coarse particles with a subset of small particles which we term impurities. Determining the amount of impurities requires information about the initial size distribution and the calculation of how this size distribution evolves in time in the supernatant and in the sedimented layer. The current paper analyses mathematically this evolution for two cases as sketched in Fig.\ref{fig1}: homogeneous sedimentation, in which the particles are mixed throughout the vial \citep{khan2012size,backes2016production}, and band sedimentation, in which the particles are initially deposited in a small slab at the top of the clear fluid \citep{backes2014edge,fadil2017chemically,gao2018lateral}. We will see that band sedimentation holds significant promise for size fractionation.  From a mathematical point of view, developing a centrifugation protocol is similar to developing an algorithm, i.e. a step-by-step sequence of instructions to obtain a certain quantifiable outcome. This paper  lays the basis for a rational, step-by-step algorithm to size fractionate a dilute polydispersed suspension of particles for the case in which the initial size distribution is known. The idea is that, to isolate a certain interval of this size distribution, we can eliminate the tails of the probability distribution in successive steps (see Fig.\ref{fig2} for an illustration of the concept). In reality the “cut” will not be sharp, so it is important to quantify the amount of impurities (error) made in each step. 

The main assumption of the paper is that the suspension is dilute. If the suspension is not dilute, the calculation of the time dependence of the particle size distribution can not be done exactly. Furthermore, the dilute case enables to establish a theoretical framework that is useful both for practical initial calculations and for further theoretical work.  We further assume the initial particle size distribution follows a log-normal. This assumption is not very restrictive. Log-normal distributions are almost universal when the particle size distribution is the result of repetitive break-up processes \citep{hosoda2011origin} (for example, the size of the nanosheets produced by liquid-phase exfoliation follows a log-normal distribution \citep{backes2014edge,khan2012size}). {Log-normal distribution is also widely used to describe particle size because it fits well for the measured distributions of many types of particles \citep{thomas1987determination} (e.g. aerosols \citep{heintzenberg1994properties}, ultrafine metal particles \citep{granqvist1976log}, soil particles \citep{dexter1972packing}) and the number, weight and area distributions of the particles are all log-normal with same standard deviation \citep{allen2013particle}.} 

A review of mathematical models of centrifugation is given in the following section, where key elements of novelty of the current analysis are highlighted. The mathematical model used in the current investigation is described in Sec.\ref{model}, and results discussing solutions of this model for homogeneous or band sedimentation are discussed in Sec.\ref{results}. To illustrate the practical application of the theory, in Sec. \ref{sec4} we describe two applied cases, fractionation of spherical metallic nanoparticles and fractionation of graphene platelets. We chose graphene because the development of a protocol to fractionate graphene would enable to overcome one of the biggest bottlenecks in large-scale graphene production, namely the large polydispersity in lateral size and thickness of the platelets produced \citep{backes2020production}. Finally, conclusions are drawn in Sec.\ref{sec5}.

\section{Overview of mathematical models of sedimentation/centrifugation}

The evolution of the particle size distribution following the application of a constant body force to a particulate system can be modeled via a set of one-dimensional transport equations, each governing the particle number density (or volume fraction) corresponding to each size class. \citet{kynch1952theory} analysed a monodisperse particulate system via one such model, excluding particle diffusion. Kynch’s model could capture the discontinuity in the volume faction profile in correspondence to the liquid-suspension and suspension-sediment interfaces observed in experiments. By extending Kynch’s theory by adding a diffusive term to the transport equation, \citet{davis1989asymptotic} analysed the diffusive broadening of the concentration fronts due to Brownian motion. \citet{blanchette2005particle} studied the evolution of particle volume fraction during sedimentation of monodisperse particles in a density stratified medium. \citet{antonopoulou2018numerical} developed a one-dimensional continuum model for sedimentation of monodisperse colloidal particles in centrifugation. To account for the fact that the sediment layer grows in size when it is sufficiently packed, they incorporated into the model an effective maximum volume fraction, derived by considering the minimum separation between two spherical colloids.

Theoretical studies on systems of 2 or 3 size classes are available \citep{al1992sedimentation,watson2005sedimentation,dorrell2010sedimentation}, whereas theoretical studies on sedimentation of widely polydispersed systems (i.e. number of classes significantly larger than 3) are scarce. \citet{esipov1995coupled} analysed a theoretical model for polydispersed suspensions. The settling velocity of each class was closed in terms of a weighted average of the volume fraction of each particle class, leading to a set of coupled Burgers equations. The initial Gaussian particle size distribution was discretized with 26 particle classes. \citet{xue2003modeling} compared experimental results with the solution of a one-dimensional convection equation (Brownian motion and hydrodynamic diffusion were neglected), considering 35 size classes. A similar model was used by \citet{abeynaike2012experimental}, with up to 8 size classes.

In contrast to the works above, we consider a fine discretization of the particle size distribution with 1000 size classes, to predict the evolution of a continuous log-normal size distribution with specified mean and variance values. A further key difference with previous works is that, we focus on the time evolution of the particle size distribution in the supernatant vs. that in the sedimented layer, while previous works focused on the time evolution of concentration profiles for each particle class. A key novelty of the current paper is that the work published so far considered single-step sedimentation, meaning that starting from a mixed suspension the simulations ended when all the particles reached the bottom of the container. In contrast, we consider multi-step sedimentation in which the sediment is removed after a given time and the remaining suspension is subjected to a further sedimentation step.

\section{Mathematical model of this work}\label{model}
We consider a dispersed system contained in a straight vial of height $H$. The body force acting on the dispersed phase (which is equal to the gravity force in sedimentation and to the centrifugal force in centrifugation) is assumed to be constant along the vial and directed parallel to the side walls of the vial (see Fig.\ref{fig1}). We seek to describe the number density $N(x,t,q)$ of particles at position $x$ and time $t$ having settling velocity $q$, with $x=0$ corresponding to the free surface of the liquid (the axis $x$ is directed towards the bottom of the vial).   We focus initially on particle velocity, and not on particle size, because calculating the time evolution of the size distribution requires stipulating a relation between size and velocity, and this relation depends on the specific shape of the particles, hence it is not universal. Once the problem is understood and modeled in the velocity space, the results can be translated in terms of size (see Sec. \ref{sec4}). 

The time evolution of $N(x,t,q)$ is governed by a convection-diffusion equation \citep{buzzaccaro2008kinetics}:
\begin{equation}
    \label{eq1}
    \frac{\partial}{\partial{t}}N(x,t,q)+\frac{\partial}{\partial{x}}(qN(x,t,q))=\frac{\partial{F_D}}{\partial{x}}.
\end{equation}
Here $F_D$ is a diffusive flux accounting for the diffusive transport of the particles in the $x$ direction (by either Brownian motion or hydrodynamic diffusion). The second term on the left-hand side represents the convection of the dispersed phase by the body force. The settling velocity $q$ is in general a function of the particle size, particle density, effective density of the suspension, concentration of the particles with velocity $q$, and concentration of the particles with velocity different from $q$ \citep{buzzaccaro2008kinetics,batchelor1982sedimentation}. The diffusive flux also depends on full size distribution. Many papers considered the problem of closing the hindered settling function for polydispersed systems \citep{uttinger2021probing,batchelor1982sedimentation,davis1994hindered,kumar2000new,abbas2006dynamics}. Because of the complexity of the multi-step situation and to produce a theoretical baseline for future work, in our simulation we do not consider hindered settling effects. We assume that the suspension is extremely dilute, i.e. the total volume fraction of the dispersed phase $\varphi_T \ll 1$, so that hydrodynamic and contact interactions between the dispersed phase elements can be neglected. In this case, $q$ becomes independent of $N$ and can be taken out of the differentiation. Furthermore, if hydrodynamic and contact interactions are negligible, the diffusive flux is also negligible. With the further assumption that Brownian motion is negligible (high P\'eclet number), Eq.\ref{eq1} simplifies to the following linear convective equation:
\begin{equation}
    \label{eq2}
    \frac{\partial}{\partial{t}}N(x,t,q)+q\frac{\partial}{\partial{x}}N(x,t,q)=0.
\end{equation}

\begin{figure}[ht]
    \centering
    \includegraphics[width=0.8\textwidth]{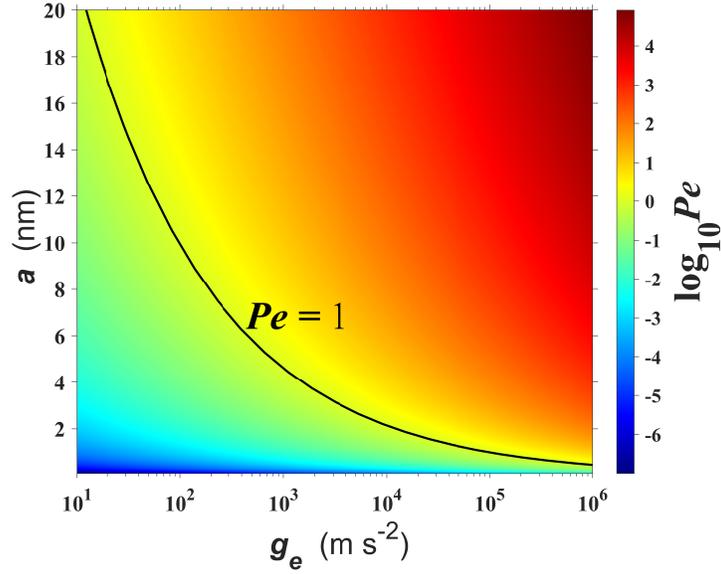}
    \caption{Color map of logarithm value of P\'eclet number of spherical particles in the ($a$, $g_e$) space. The black line is $Pe = 1$.}
    \label{fig3}
\end{figure}

The characteristic time scale for convection over a length $H$ is $H⁄q$. The ratio of the diffusive time scale $H^2/D$ to the convective time scale is the P\'eclet number $Pe=Hq/D$, where $D$ is the diffusion coefficient of the particle \citep{mewis2012colloidal}.  
For spherical particles, the condition for particle radius $a$ of negligible diffusion in centrifugation ($Pe>1$) corresponds to $a^3 > 3k_B T/4 \pi H (\rho_p - \rho_l) g_e $, where $k_B T$ is the characteristic thermal energy, $\rho_p$ and $\rho_l$ as densities of particles and solvent respectively, $g_e$ as the centrifugal acceleration, and the relation of $a$ and $g_e$ making $Pe=1$ is plotted in Fig.\ref{fig3} assuming $H = \SI{1}{cm}$, $(\rho_p - \rho_l)= \SI{1000}{kg\per\meter^3}$, and at room temperature. 

We solve equation \ref{eq2} numerically and analytically for the initial condition:
\begin{equation}
    \label{eq3}
    N(x,t=0,q)=N_0(x,q),
\end{equation}
where $N_0(x,q)$ is the initial particle number density of particles with velocity $q$ at position $x$. We consider two situations: the particles are homogeneously distributed in a slab of height $h<H$ near the top of the vial (case 1, band sedimentation); or the particles are homogeneously distributed in $x \in [0,H)$ (case 2, homogeneous sedimentation). The model equation is solved in a half-bounded region $x \in [0, \infty)$, where $x=0$ corresponds to the free interface at the top of the vial. The particles are considered to be sedimented when they reach $x=H$.

In the dilute limit, the settling velocity is only a function of the particle geometry. The settling velocity $q$ of an isotropic (spherical) particle (Eq.\ref{eq8}) or an anisotropic (disk) particle (Eq.\ref{eq9}) is given in Sec.\ref{sec4}. We consider an initial log-normal distribution of $q$ as an illustration of the application of the theory (note that for disks the settling velocity is log-normally distributed if lateral size $d$ and thickness $L$ are both log-normally distributed and independent \citep{larsen2005introduction}).

For $n_q$ discrete size species Eq.\ref{eq2} becomes
\begin{equation}
    \label{eq4}
    \frac{\partial}{\partial{t}}N_i(x,t;q_i)+q_i\frac{\partial}{\partial{x}}N_i(x,t;q_i)=0,\quad  i=1,2,\cdots,n_q,
\end{equation}
where $N_i(x,t;q_i)$ is the number density of species $i$ with settling velocity $q_i$. The initial condition for species $i$ is
\begin{equation}
    \label{eq5}
    N_i(x,0;q_i)=f_0(q_i)n_0(x),
\end{equation}
where $n_0(x)$ is the initial total particle number density including all the species at position $x$, and $f_0(q_i)$ is the frequency of species $i$ in the original sample. For case 1 (band sedimentation), we specify
\begin{equation}
    \label{eq6}
    n_{0}^{*}(x^*)=\begin{cases}
    1 & \text{for $0 \leq x^* \leq 0.1$} \\
    0 & \text{for $0.1 < x^* \leq 1$}
    \end{cases}
    .
\end{equation}
For case 2 (homogeneous sedimentation), we specify
\begin{equation}
    \label{eq7}
    n_{0}^{*}(x^*)=1,\quad 0 \leq x^* \leq 1.
\end{equation}
Here variables with superscript ($^*$) are dimensionless, and $x^*=x/H$, $n_{0}^{*}=n_0/m$ where $m$ is the initial number density value everywhere in the particle-laden region. 

The numerical procedure to solve Eq.\ref{eq4} proceeds as follows. From the probability distribution function (p.d.f.) of $q$, $p(q)$, a discrete frequency distribution $f(q_i)$ can be calculated for $i=1,\cdots,n_q$, dividing the range $[0,Q]$ into $n_q$ slabs of size $\Delta q=Q/n_q$. Here $Q$ is an assigned maximum value of $q$ at which the upper tail of the p.d.f. is cut. The value of  $f(q_i)$ gives the percentage fraction of particles with velocities between $q_i$ and $q_i+\Delta q$, and is calculated as  $f(q_i)=p(q_i)/\sum_{i=1}^{n_q}p(q_i)$. In our discretization, the nodes are centered at each slab, so $q_i=(i-0.5)\Delta q$. The frequency distribution at time $t=0$ is denoted as $f_0(q_i)$. Eq.\ref{eq4} is then solved by a finite difference method in which spatial derivative is discretized by a first-order upwind scheme and time is integrated by a first-order Euler scheme.

The solution of Eq.\ref{eq4} gives the probability of having particles of a given class $q_i$ for any position $x$ at time $t$. In a multistep sedimentation process, the interest is in characterizing how the particle distribution in the supernatant differs from that in the sedimented layer. The number of particles of size $q_i$ in the supernatant, denoted as $n^s (q_i)$ is calculated by integrating over the entire length of the vial, assuming that the particles which have crossed the boundary $x=H$ as the sedimented layer: $n^s (q_i)= \int_0^H N(t,x;q_i) \rm{d} x$. The number of particles in the sedimented layer is $n^b (q_i)= \int_H^{\infty} N(t,x;q_i) \rm{d} x$. The corresponding frequencies are $f^s (q_i)=n^s (q_i)/\sum n^s (q_i)$  and $f^b (q_i)=n^b (q_i)/\sum n^b (q_i)$. The sum $n^t$  of the number of particles in the supernatant and in the sedimented layer is evidently equal to the initial number of particles: $n^t=n^s+n^b=\int_0^H n_0 (x) \rm{d} x$.

\section{Results and discussion}\label{results}
\subsection{Time evolution of the frequency distribution: supernatant v.s. sediment layer}\label{sec3.1}
\begin{figure}
    \centering
    \includegraphics[width=1\textwidth]{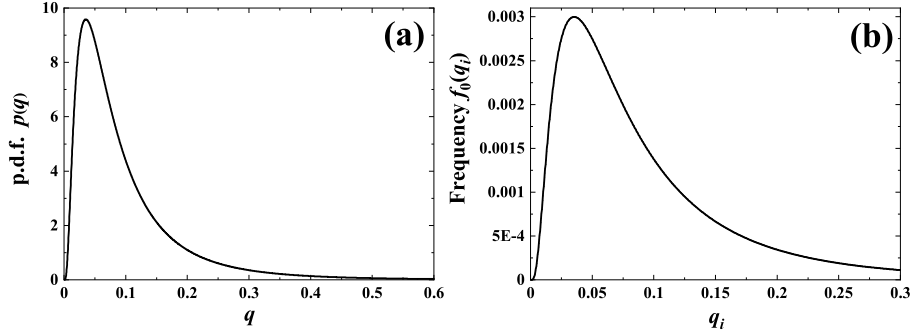}
    \caption{(a) p.d.f. of a log-normal distribution with mean value 0.1 and variance 0.01. (b) Discretized frequency distribution for 1000 species corresponding to (a).}
    \label{fig4}
\end{figure}
\begin{figure}[ht]
    \centering
    \includegraphics[width=1\textwidth]{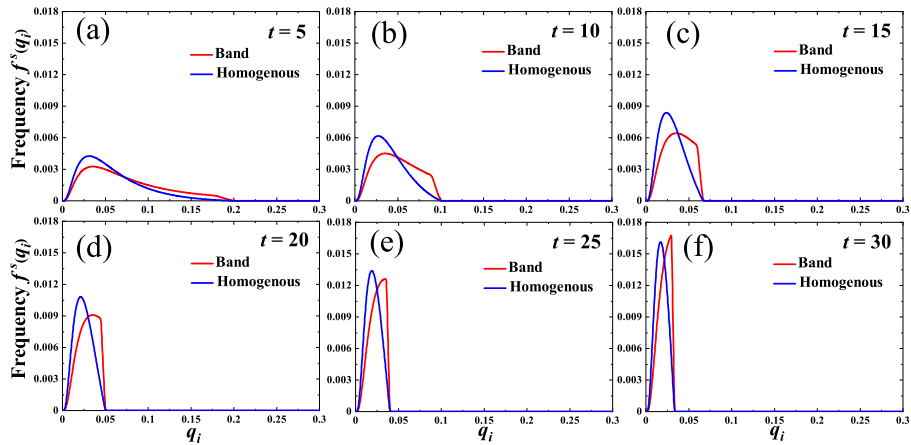}
    \caption{Time evolution of the frequency distributions in the supernatant for case 1 (red lines) and case 2 (blue lines).}
    \label{fig5}
\end{figure}
\begin{figure}[htbp]
    \centering
    \includegraphics[width=1\textwidth]{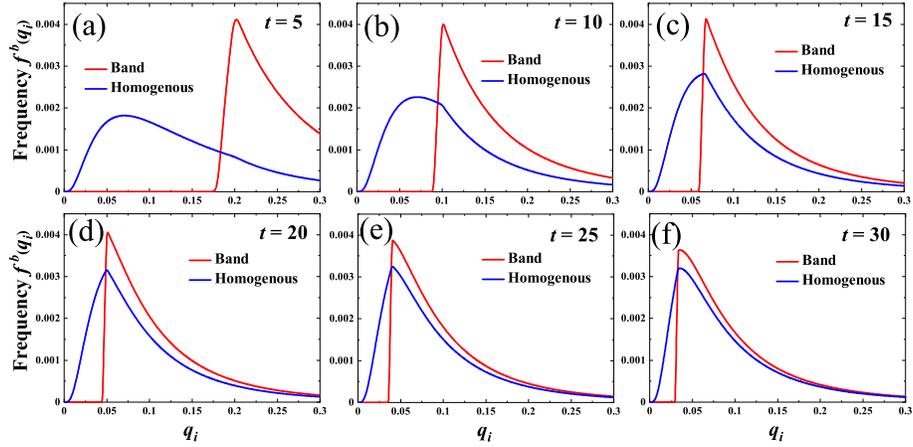}
    \caption{Time evolution of the frequency distributions in the sediment layer for case 1 (red lines) and case 2 (blue lines).}
    \label{fig6}
\end{figure}
\begin{figure}[htbp]
    \centering
    \includegraphics[width=1\textwidth]{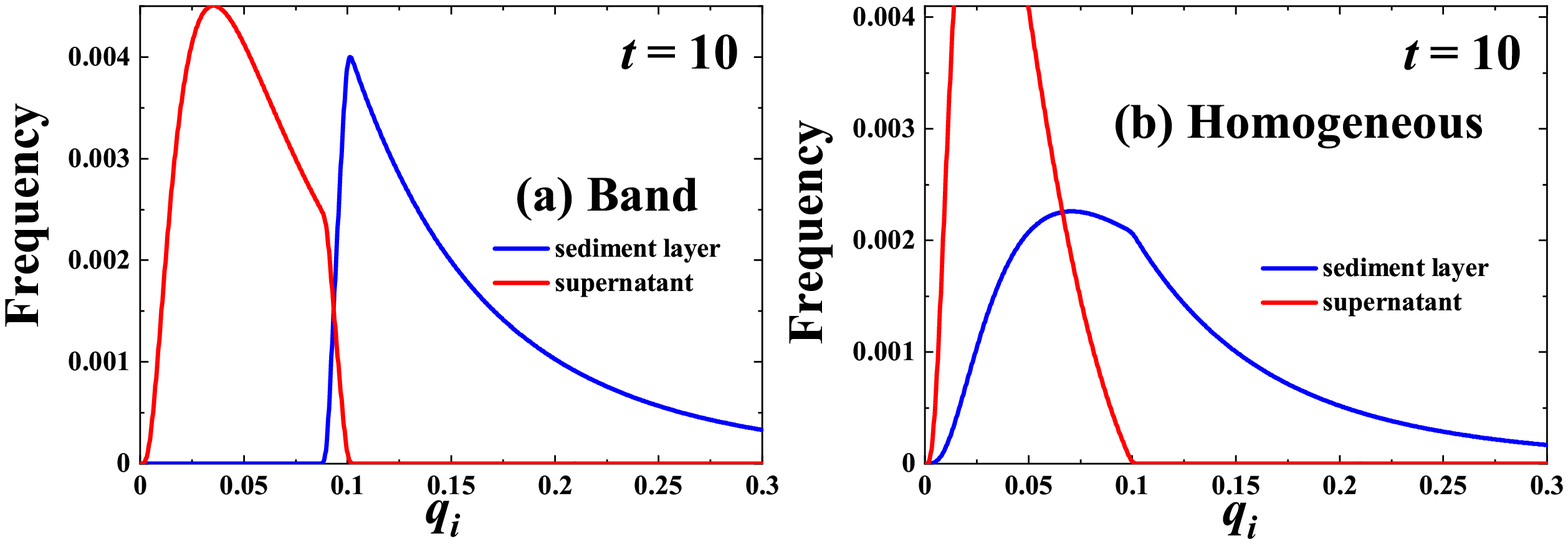}
    \caption{Frequency distributions of particles in the supernatant (red lines) v.s. in the sediment layer (blue lines) at $t=10$ for (a) case 1 band sedimentation, and (b) case 2 homogeneous sedimentation.}
    \label{fig7}
\end{figure}
A log-normal p.d.f. $p(q)$ with mean value of $q$ as 0.1 and variance of $q$ as 0.01 is shown in Fig.\ref{fig4}(a). Note that the settling velocity $q$ is non-dimensionalized by a characteristic velocity (e.g. the settling velocity of a particle with certain size). It is seen that for these parameters most of the distribution falls within $q \in (0,0.3)$, so we choose the maximum size $Q = 0.3$. By choosing $Q = 0.3$ and $n_q = 1000$, the interval from $q = 0$ to $0.3$ is discretized into 1000 slabs of increment $\Delta q = 0.0003$. The corresponding frequency distribution $f_0(q_i)$ is shown in Fig.\ref{fig4}(b).

Time evolutions of the frequency distributions in the supernatant and in the sediment layer are shown in Fig.\ref{fig5} and Fig.\ref{fig6}, respectively, for both case 1 (band sedimentation) and case 2 (homogeneous sedimentation). From Fig.\ref{fig5}, it is seen that there are fewer size classes in the supernatant as time goes by because the larger particles sediment on the bottom quicker. The largest size classes in the supernatant are the same for band and homogeneous sedimentation at the same time. However, the moving front of the frequency distribution curve of the supernatant in band sedimentation is sharper than that in homogeneous sedimentation. From Fig.\ref{fig6}, it is seen that the sediment layer in homogeneous sedimentation always contains all the size classes, whereas in band sedimentation the smaller size classes come to the sediment layer at later times. In Fig.\ref{fig7}, the frequency distributions in the supernatant and in the sediment layer at the same time $t=10$ are plotted together for both band and homogeneous sedimentation. It is seen that the curves overlap slightly in band sedimentation, which means there is clear distinction between the size classes in the supernatant and the size classes in the sediment layer. However, all the size classes in the supernatant appear in the sediment layer for homogeneous sedimentation.      
\subsection{Particle fractions in the supernatant and in the sediment layer}\label{sec3.2}
\begin{figure}
    \centering
    \includegraphics[width=0.6\textwidth]{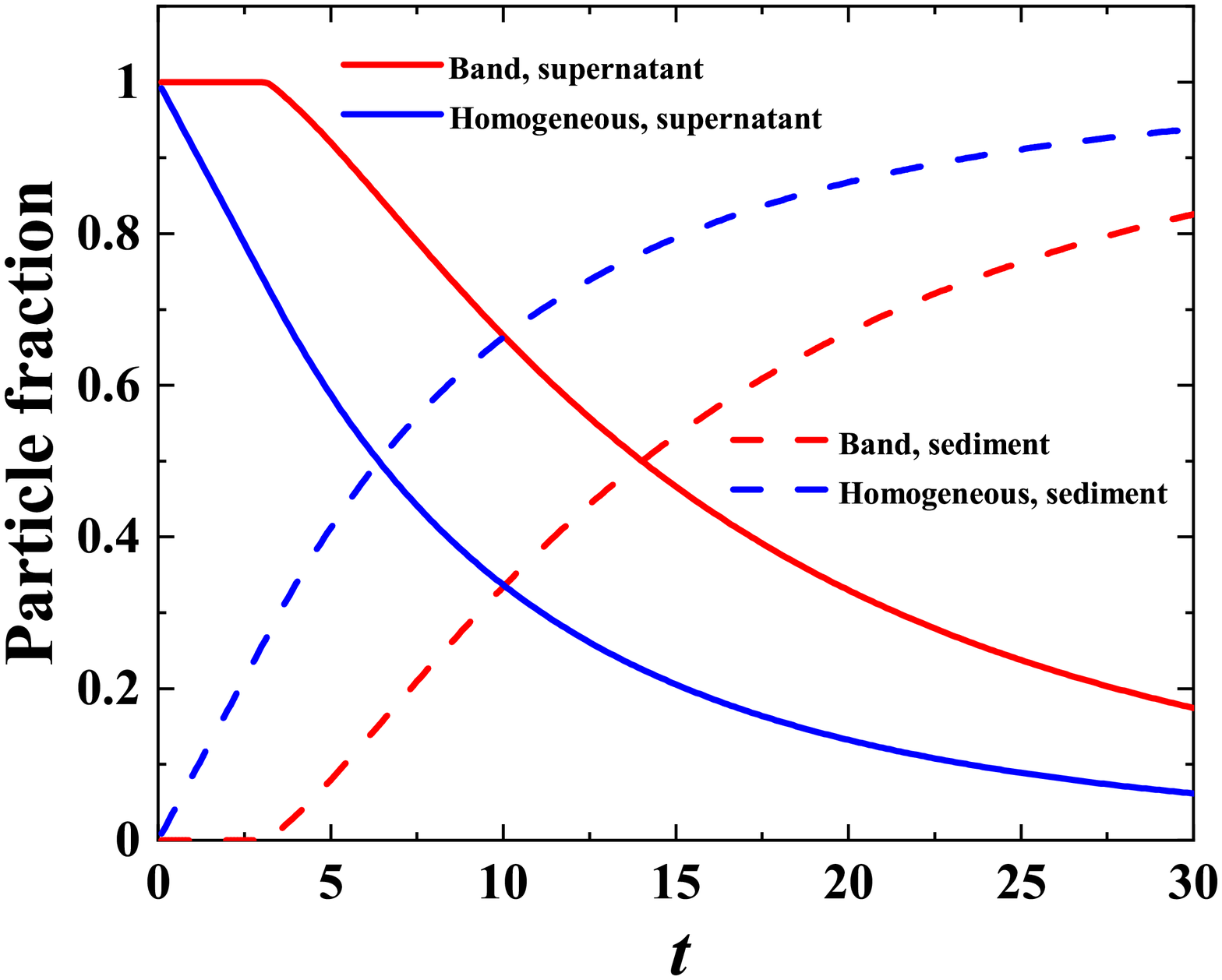}
    \caption{Time evolution of particle fractions in the supernatant and in the sedimented layer for both band and homogeneous sedimentation.}
    \label{fig8}
\end{figure}
\begin{figure}
    \centering
    \includegraphics[width=0.6\textwidth]{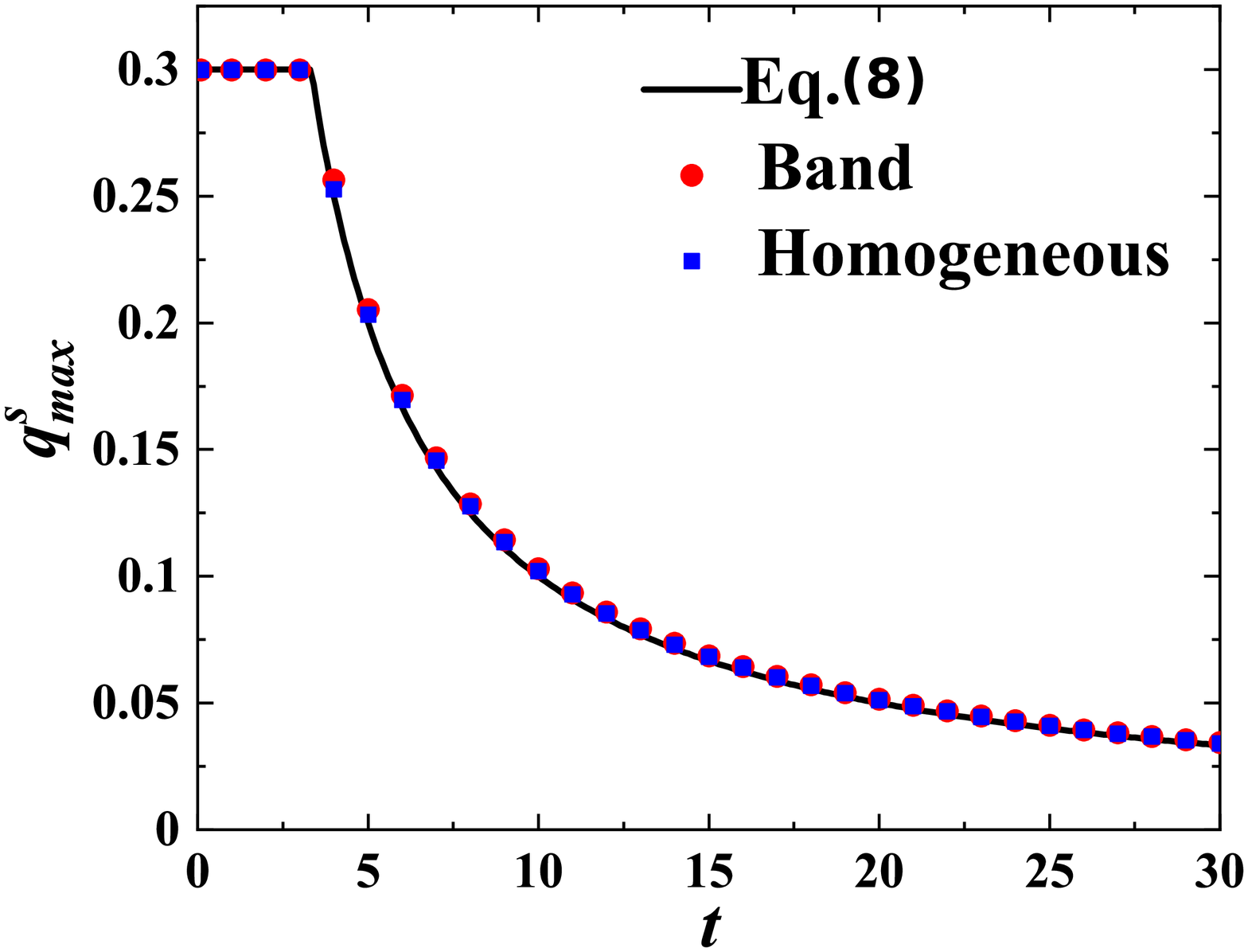}
    \caption{The value of $q_{max}^s$ versus time $t$. Eq.(8), band and homogeneous sedimentation are shown by the solid line, circular and square symbols, respectively.}
    \label{fig9}
\end{figure}

Fig.\ref{fig8} shows the time evolution of particle fractions in the supernatant and in the sedimented layer, respectively, for both band and homogeneous sedimentation. Due to the existence of clear fluid under the particle-laden layer in band sedimentation, the decreasing of particle fraction in the supernatant is delayed, whereas the particle fraction in the supernatant starts to decrease from the beginning in homogeneous sedimentation. Moreover, it is seen that particle fraction in the supernatant decays faster in homogeneous sedimentation than that in band sedimentation by comparing the slopes of the curves of the two cases.

In both band and homogeneous sedimentation the initial frequency distribution is “cut” by a front that moves to the left with a certain velocity (see Fig.\ref{fig5}). This velocity can well be characterized as the velocity of $q_{max}^s$, the largest particle size in the supernatant. Because the transport equation Eq.\ref{eq4} is linear, $N_i (x,t;q_i)$ can be easily calculated analytically. From the value of $N_i (x,t;q_i)$, particle number of species $q_i$ in the supernatant at time $t$, which is $n^s (q_i,t)$, can be calculated by integration. Setting $n^s (q_i,t)=0$ provides the following relationship between $q_{max}^s$ and $t$ (see appendix for details):
\begin{equation}
    \label{eq11}
    q_{max}^s=\begin{cases}
    H/t & \text{for $t > H/Q$} \\
    Q & \text{for $t \leq H/Q$}
    \end{cases}
    ,
\end{equation}
where $H$ is the length from the free interface to the bottom of the vial, and $Q$ is the largest value of $q$ of the original sample. The validity of this expression is demonstrated in Fig.\ref{fig9}, where the simulated values of $q_{max}^s$ with varying time are plotted.

\subsection{Minimisation of impurities}\label{sec3.3}
The presence of a front that moves with velocity given by Eq.\ref{eq11} suggests a protocol to isolate a certain interval of the frequency distribution. Suppose we would like to separate particles with $q \in (q_{min},q_{max} )$ from all the other particles, where $q_{min}$ and $q_{max}$ are somewhere in the middle of the frequency distribution.  An intuitive protocol is to centrifuge the suspension for a certain time and discard the sediment. This would give in the supernatant a suspension containing only particles with velocities less than a certain threshold. Then we could take the supernatant, centrifuge it again and in this case remove the supernatant. In the sedimented layer, we would have particles having an intermediate range of velocities larger than a minimum threshold and smaller than a maximum threshold. The sequence could be repeated again, with different centrifugation times assigned to each step. The problem is to predict how the {\it observed} thresholds depend on the {\it assigned} $q_{min}$ and $q_{max}$ for a given sequence of centrifugation times. Solving this problem requires choosing the centrifugation time rationally so that the amount of impurities is minimized. We have found in the previous section that the initial size distribution cannot be cut sharply into two parts (i.e. the frequency curves in the supernatant and in the sediment layer overlap at each time), and a certain amount of small velocity (finer) particles will always reach the bottom.

An insight from the previous section is that the time scale of motion of the front is given by the ratio of the vial height $H$ and the front velocity. This suggests that the centrifugation times should be chosen based on the ratio of $H$ with $q_{max}$ and $q_{min}$. We therefore propose the following algorithm, whose convergence and error we aim to characterize (in the same way as done for algorithms implementing numerical methods in scientific computing):

\begin{hangparas}{.25in}{1}
\textbf{Step 0}: Calculate centrifugation times $t_{min}=H/q_{max}$ and $t_{max}=H/q_{min}$, and set $\Delta t=(t_{max}-t_{min})/(\mathcal{N}-1)$ based on the total centrifugation steps $\mathcal{N}$.

\textbf{Step 1}: Centrifuge the suspension, using a constant body force, for a time $t_1=t_{min}$. Move the supernatant to another test tube which is to be centrifuged in the next step. Discard the sediment.

\textbf{Step 2}: Centrifuge the supernatant from the previous step 1 for a time $t_2=t_1+\Delta t$. Move the supernatant to another test tube for the next step centrifugation. Collect the sediment as part of the separated sample.

\textbf{Step 3}: Centrifuge the supernatant from the previous step 2 for time $t_3=t_2+\Delta t$. Move the supernatant to another test tube for the next step centrifugation. Collect the sediment as part of the separated sample.

$\cdots$
 
\textbf{Step $\mathcal{N}-1$}: Centrifuge the supernatant from the previous step $\mathcal{N}-2$ for time $t_{\mathcal{N}-1}=t_{\mathcal{N}-2}+\Delta t$. Move the supernatant to another test tube for the next step centrifugation. Collect the sediment as part of the separated sample.

\textbf{Step $\mathcal{N}$}: Centrifuge the supernatant from the previous step $\mathcal{N}-1$ for time $t_\mathcal{N}=t_{max}$. Discard the supernatant. Collect the sediment as part of the separated sample.

\textbf{Step $\mathcal{N}+1$}: Assemble all the collected sediment from Steps $2,3,\cdots,\mathcal{N}-1$ and $\mathcal{N}$ as the separated sample.
\end{hangparas}

According to Eq.\ref{eq11}, the largest value of $q$ in the supernatant at the end of Step 1 is $q_{max}$. This means that all the particles with $q>q_{max}$  in the original sample have settled at the bottom wall in Step 1, and therefore the sediment should be completely discarded in Step 1 as according to the prediction it does not contain useful particles. On the other hand, according to the prediction at the end of Step $\mathcal{N}$, the largest value of $q$ in the supernatant is $q_{min}$, which means the supernatant contains particles smaller than the desired ones. Hence the supernatant at the end of step $\mathcal{N}$ should be discarded. The model predicts that particles with $q \in (q_{min},q_{max})$ have settled on the bottom wall gradually from Step 2 to Step $\mathcal{N}$. That is why the sediment layers from steps $2 \cdots \mathcal{N}$ should be collected and mixed together, and the collected sample has a distribution whose shape depends on the number of total centrifugation steps $\mathcal{N}$.
\subsubsection{Effect of the number of steps}\label{sec3.3.1}
An example is given in the current subsection to show how the multi-step approach works and how the number of centrifugation steps may be chosen. Suppose particles with $q \in (0.025,0.045)$ from the log-normally distributed system shown in Fig.\ref{fig4}(b) are desired. This interval is chosen because it contains the value of $q$ corresponding to the peak of the initial frequency distribution, $q=0.035$. The calculated values of $t_{min}$ and $t_{max}$ are 22.2 and 40, respectively (step 0). We need to devise $\mathcal{N}$ steps of successive centrifugation, for which $t_1=22.2$ in Step 1 and $t_\mathcal{N}=40$ in Step $\mathcal{N}$. We would like to know how the purity of the sample depends on $\mathcal{N}$.
\begin{figure}
    \centering
    \includegraphics[width=1\textwidth]{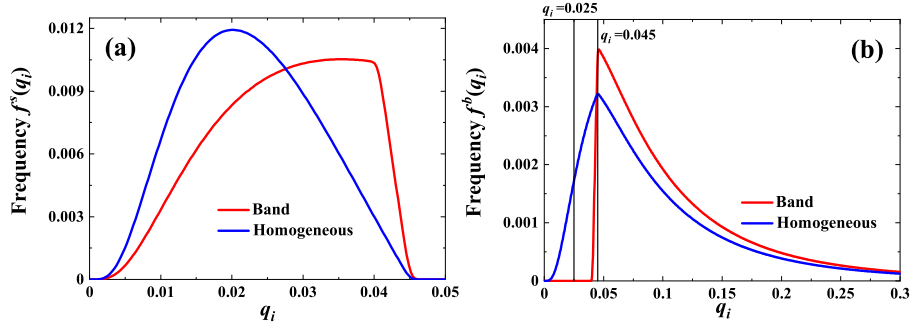}
    \caption{Frequency distributions after Step 1 of centrifugation for particles (a) in the supernatant and (b) in the sedimented layer. (Band sedimentation is shown as red lines, and homogeneous sedimentation as blue lines. The two vertical thin lines in (b) mark $q$ equals 0.025 and 0.045, respectively.)}
    \label{fig10}
\end{figure}

Because we know that band sedimentation gives a sharper division of the initial frequency distribution, we anticipate that the level of purity given by band sedimentation is larger than that by homogeneous sedimentation. Hence, we consider two cases: in case A, the initial condition for each step of centrifugation is that the particles are uniformly dispersed within a top layer of length 0.1 as band sedimentation; in case B, the initial condition of each centrifugation step is that the particles are uniformly dispersed in the whole solvent of length 1 as homogeneous sedimentation.

The frequency distributions of the particles in the supernatant and in the sediment layer after Step 1 of centrifugation are shown in Fig.\ref{fig10}. In agreement with the prediction,  Fig.\ref{fig10}(a) shows that all the particles remaining in the supernatant are of $q$ smaller than 0.045. The sediment layer at the end of Step 1 contains, as expected, particles larger than the desired ones, and also a fraction of the desired ones from the original sample as indicated by the frequency distributions between $q=0.025$ and $q=0.045$ (marked by two thin vertical lines in Fig.\ref{fig10}(b)). This fraction, i.e. the ratio of the numbers of particles with $q \in (0.025,0.045)$ in the sedimented layer after Step 1 and in the original sample, is 13.6\% for case A and 78.4\% for case B, respectively. This means the yield will be much higher in band sedimentation than that in homogeneous sedimentation.
\begin{figure}
    \centering
    \includegraphics[width=1\textwidth]{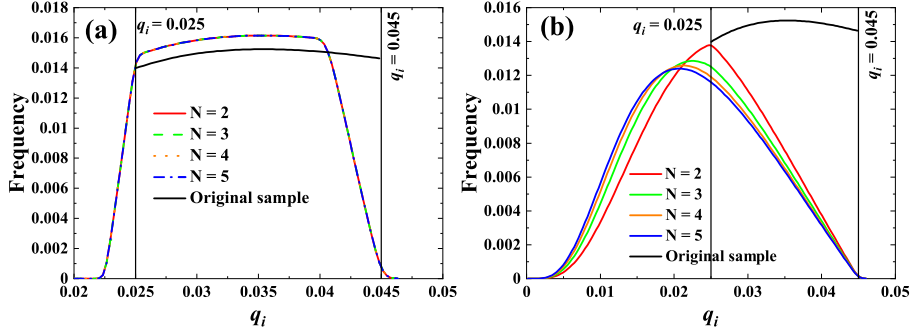}
    \caption{Frequency distributions of the separated sample as the total number of centrifugation steps $\mathcal{N}$ equals 2, 3, 4 and 5 for (a) case A (band sedimentation) and (b) case B (homogeneous sedimentation), together with that of particles within the desired range in the original sample. (The two vertical lines mark $q$ as 0.025 and 0.045 in both (a) and (b).)}
    \label{fig11}
\end{figure}

We now analyze how the shape of the frequency distribution of the separated sample changes with $\mathcal{N}$. This will give us insights into how to choose the number of centrifugation steps in the separation algorithm. Fig.\ref{fig11} shows the frequency distributions of the separated sample as $\mathcal{N}$ equals 2, 3, 4 and 5, respectively. For comparison, the frequency distribution of particles with $q \in (0.025,0.045)$ in the original sample is also shown in Fig.\ref{fig11}. It is seen that in case A (band sedimentation) the frequency distribution curve nearly remains the same as $\mathcal{N}$ increases, thus the separation algorithm converges fast and using $\mathcal{N}=2$ gives predictions that are no worse than those obtained for $\mathcal{N}=3$. There are differences with the “desired” size distribution, shown in black. In particular, while the maximum value of  $q$  is 0.045 as desired, the fraction of particles with velocities close to 0.045 is smaller than that in the original distribution (there is an evident drop in frequency for $q$ larger than about 0.042). In case B (homogeneous sedimentation), the separation algorithm instead does not converge. The curve becomes smoother as $\mathcal{N}$ increases, assuming an approximate bell shape, and the peak value shifts to the left (smaller values of $q$) as $\mathcal{N}$ increases. The shape of the frequency distribution is quite different from the desired one, although the range of $q$ covers the desired range.
\begin{figure}
    \centering
    \includegraphics[width=1\textwidth]{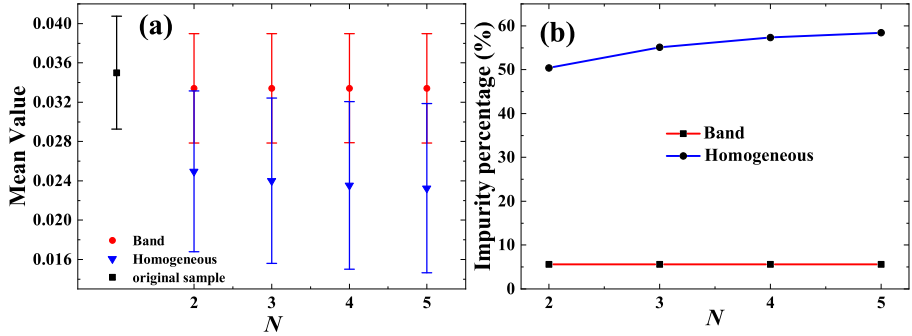}
    \caption{(a) Mean value and standard deviation of $q$ and (b) the impurity percentage of the separated sample for different $\mathcal{N}$ values for both band and homogeneous sedimentation. The corresponding values of the desired particles in original sample are also shown in (a).}
    \label{fig12}
\end{figure}

To characterize the convergence properties more synthetically, Fig.\ref{fig12}(a) shows the mean value and standard deviation of $q$ of the separated sample for $\mathcal{N}$ equals 2, 3, 4 and 5. It is seen that the mean value in case B (homogeneous sedimentation) is smaller than that in case A (band sedimentation), and standard deviation in case B is larger than that in case A. The impurity percentage of the separated sample, defined as the ratio of the number of particles with $q$ outside  (0.025,0.045) to the total particle number of the sample, is shown in Fig.\ref{fig12}(b). In case A the statistics converge fast and the impurity level is always below 10\%. In case B, the impurity level is around 50\% for $\mathcal{N}$ equals 2 and increases as the number of steps increases.



\subsubsection{An improved fractionation protocol}\label{sec3.3.2}
Based on the results from the last subsection, it is inferred that choosing $\mathcal{N}=2$ minimizes the impurity percentage. This suggests that a simple two-step approach, as follows, is practical and beneficial for the fractionation of a given range $(q_{min},q_{max})$:

\begin{hangparas}{.25in}{1}
\textbf{Step 1}: Centrifuge the suspension for time $t_1$, keep the supernatant for the next step centrifugation and discard the sediment.

\textbf{Step 2}: Centrifuge the supernatant from Step 1 for time $t_2$, then collect the sediment as the separated sample and discard the supernatant.
\end{hangparas}
\begin{figure}
    \centering
    \includegraphics[width=1\textwidth]{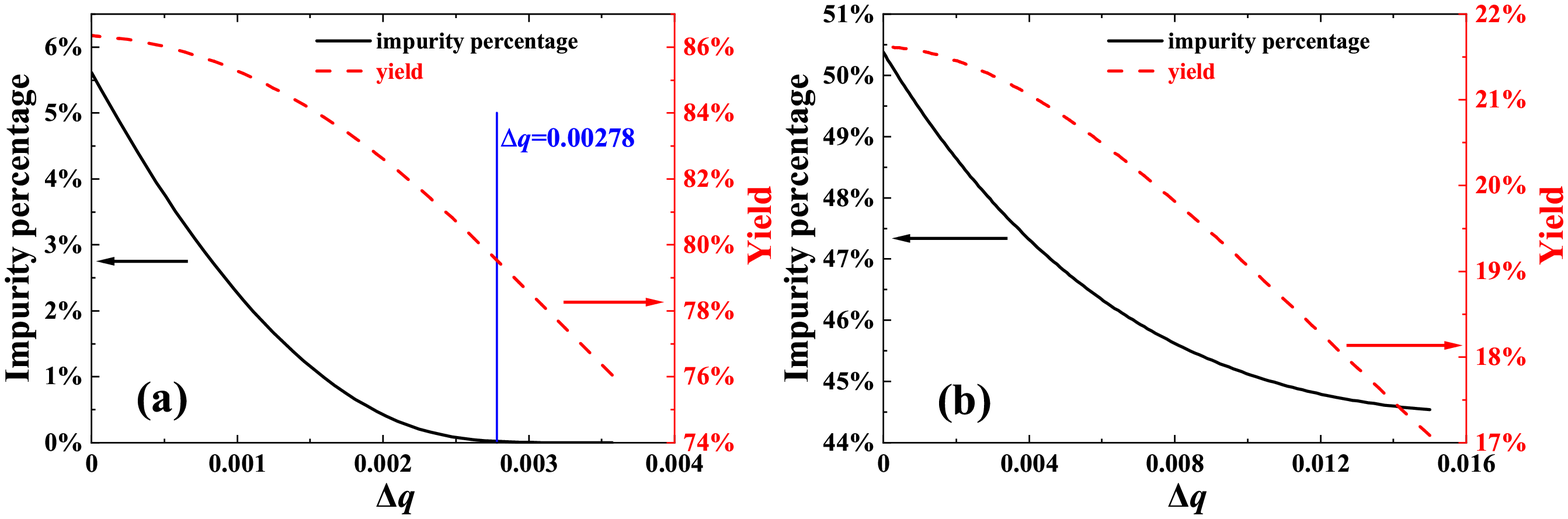}
    \caption{The impurity percentage and yield as $\Delta q$ increases for (a) case A and (b) case B, respectively. The vertical line in (a) marks $\Delta q$ equals 0.00278.}
    \label{fig13}
\end{figure}

\begin{figure}
    \centering
    \includegraphics[width=1\textwidth]{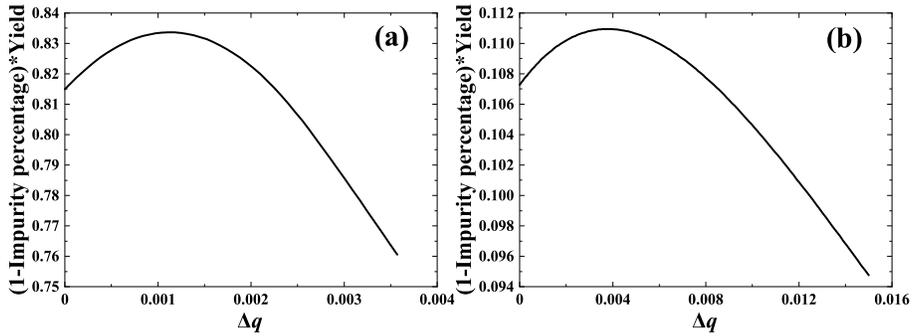}
    \caption{The value of (1-impurity percentage)$\times$yield as $\Delta q$ increases for (a) case A and (b) case B, respectively.}
    \label{fig14}
\end{figure}

There is clearly an arbitrariness in the choice of $t_1$ and $t_2$. The previous subsection suggests that the time scales should be $t_1=H/q_{max}$ and $t_2=H/q_{min}$, but choosing values that are close, but not identical, to these two limits may give better result. The impurities in the separated sample are particles with $q$ smaller than $q_{min}$. Thus, the time $t_2$ could be decreased to let fewer smaller particles sediment on the bottom wall and decrease the impurity percentage. To explore this possibility, we set $t_2=H⁄(q_{min}+\Delta q)$ and study the fractionation efficiency as a function of the parameter $\Delta q$ ($0<\Delta q<(q_{max}-q_{min})$). Apart from the impurity percentage, the other quantity of interest is the yield, defined as:
\begin{equation*}
    \textrm{yield}=\frac{\textrm{number of particles with $q \in (q_{min},q_{max})$ in the separated sample}}{\textrm{number of particles with $q \in (q_{min},q_{max})$ in the original sample}}.
\end{equation*}
By increasing $\Delta q$, we expect a decrease in impurity level but also a decrease in yield. To see the effect of increasing $\Delta q$, cases A and B in the last subsection are considered here following the modified two-step approach described above. The simulation results, shown in Fig.\ref{fig13}, confirm that the impurity percentage and yield both decrease as $\Delta q$ increases. An optimised algorithm would have a comparatively large yield and a large purity. Therefore we use an objective function, (1-impurity percentage)$\times$yield, to indicate the balance between yield and purity, as shown in Fig.\ref{fig14}. This objective function has a peak value, which corresponds to $\Delta q \simeq 0.001$ and $\Delta q \simeq 0.004$, for cases A and B respectively. The fact that  $\Delta q \ll (q_{max}-q_{min})$ means that our original choice of $\Delta q=0$ was actually quite close to the optimal. Looking at Fig.\ref{fig13}, however, one can see that while the yield does not decrease much by choosing the optimal value instead of $\Delta q=0$, the impurity percentage drops substantially in case A (from 5.5\% to about 1.5\%) because the slope of the impurity percentage curve is large when $\Delta q$ is small. This example illustrates that a finely tuned algorithm, or fractionation protocol, can have substantial effects on the purity level. Another interesting modification is to replace $t_2=H/q_{min}$ with $t_2=(H-h)/q_{min}$ in case A, where $h$ is the initial length of the particle-laden region, and this is equivalent to choose $\Delta q=q_{min} (h⁄(H-h))$. With this choice the predicted impurity percentage is 0 in case A. As indicated by Fig.\ref{fig13}(a), the impurity percentage is 0 when $\Delta q=0.025 \times 0.1⁄(1-0.1)=0.00278$.

\section{Practical applications of the theory}\label{sec4}

\subsection{Fractionation of {nearly spherical metallic nanoparticles} (isotropic particles)} \label{sec:metallic}

Nominally spherical metallic nanoparticles with well-controlled size range are important for many applications, ranging from drug delivery \citep{pankhurst2003applications}, medical diagnostics and sensors \citep{alanazi2010biopharmaceutical}, electrochemical applications \citep{pingarron2008gold} to catalytic applications \citep{hennebel2012microbial}. For spherical particle of radius $a$ in Stokes flow, the single-particle settling velocity is \citep{guazzelli2011fluctuations}
\begin{equation}
    \label{eq8}
    q=\frac{2}{9}\frac{(\rho_p - \rho_l)}{\mu} g_e a^2.
\end{equation}
Here $\rho_p$ and $\rho_l$ are densities of the particle and medium liquid, respectively, $\mu$ is the dynamic viscosity of the liquid, and $g_e$ is the equivalent $g$-force of the centrifugation $g_e = R\omega^2$ where $R$ is the rotor radius and $\omega$ is the rotational speed of the rotor.
\begin{figure}
    \centering
    \includegraphics[width=0.8\textwidth]{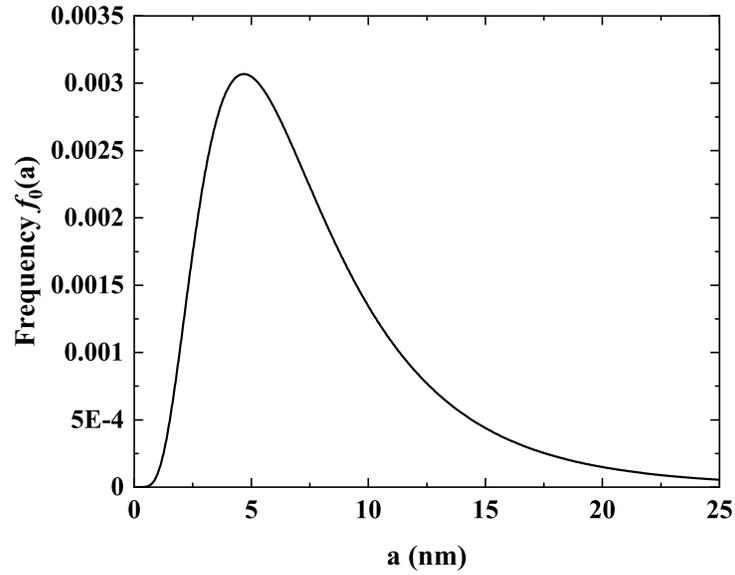}
    \caption{Radius frequency distribution of the chosen AuNP sample.}
    \label{fig15}
\end{figure}
\begin{figure}
    \centering
    \includegraphics[width=0.8\textwidth]{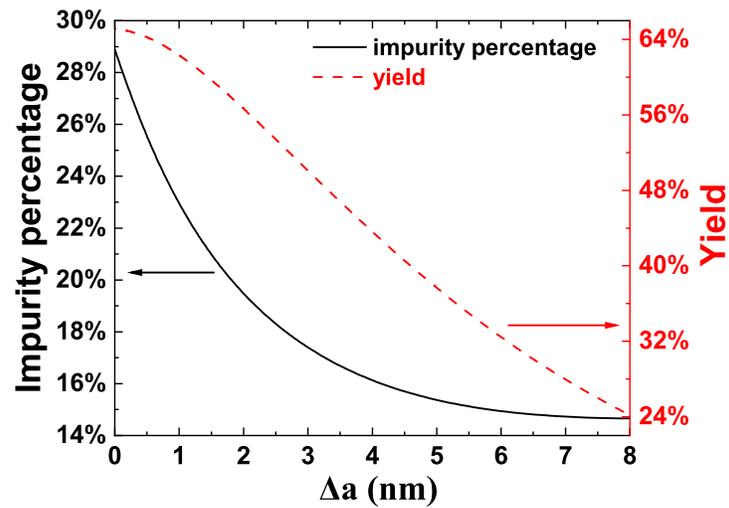}
    \caption{Effect of the offset value on impurity percentage and yield for homogeneous centrifugation.}
    \label{fig16}
\end{figure}
\begin{figure}
    \centering
    \includegraphics[width=0.8\textwidth]{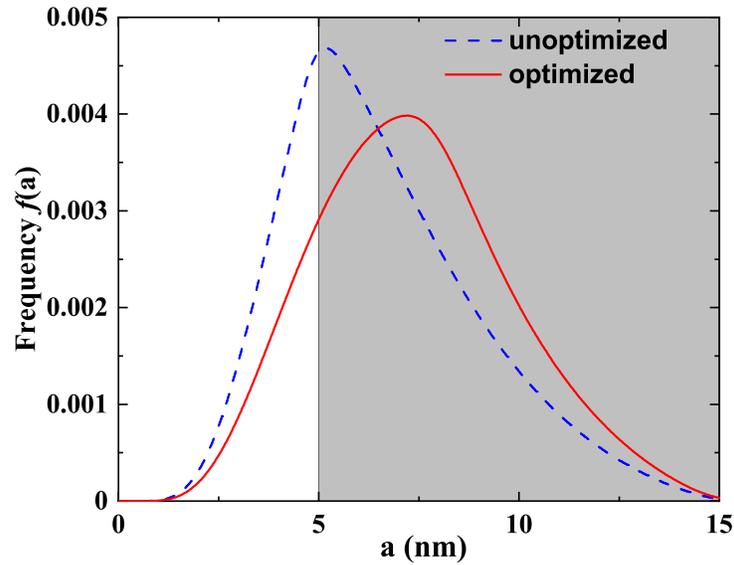}
    \caption{Size frequency distributions of the fractionated AuNPs by the unoptimized or optimized (choosing $\Delta a$ as 3 nm) two-step procedures, shaded region is the desired size range (5-15 nm).} 
    \label{fig17}
\end{figure}
\begin{figure}
    \centering
    \includegraphics[width=1.0\textwidth]{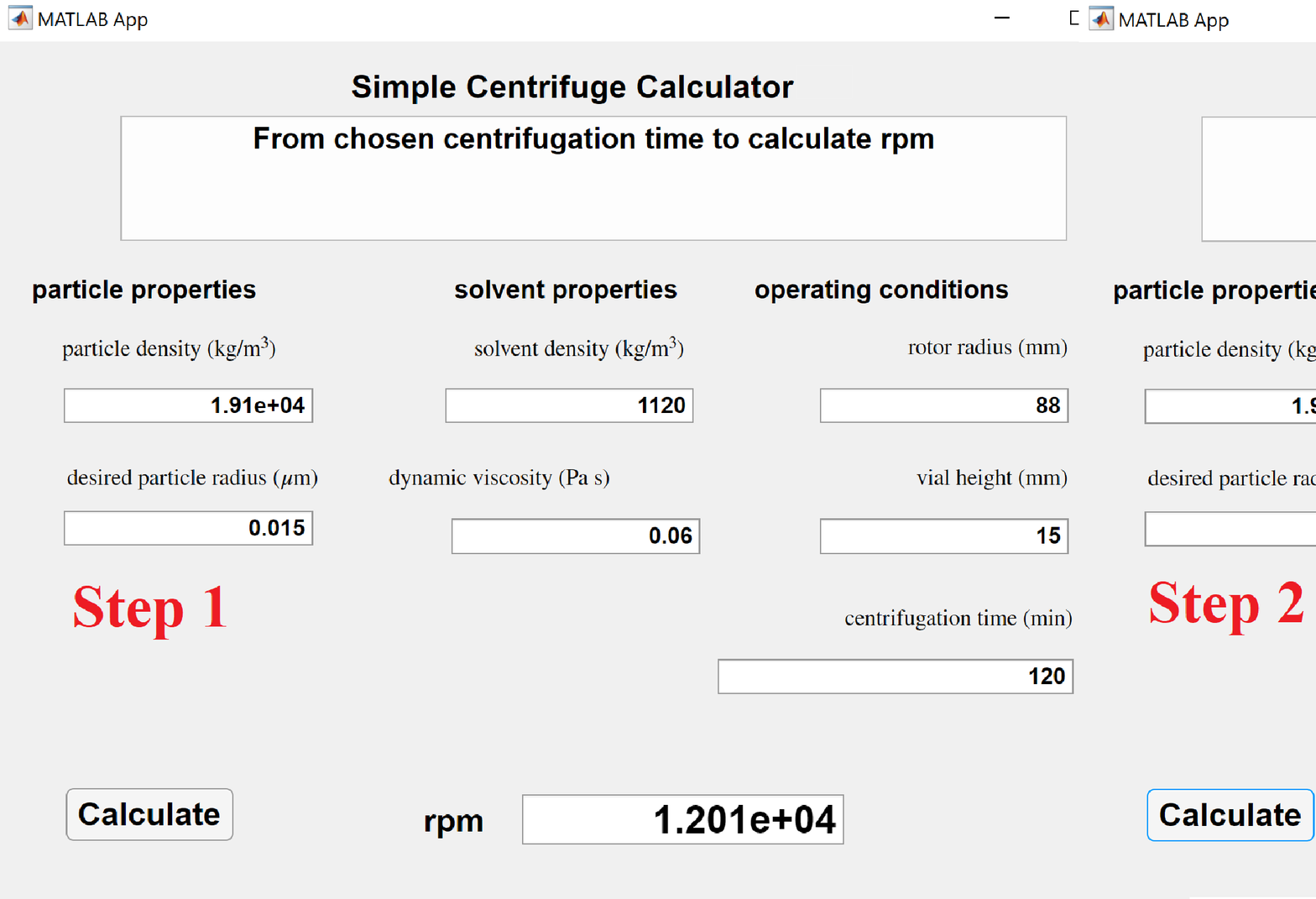}
    \caption{Values of the rpms for the optimized two-step procedure given by the centrifuge calculator.}
    \label{fig18}
\end{figure}

To illustrate how to predict the impurity percentage and yield during a centrifugation process and to design an optimal two-step procedure aiming at lower impurity percentage, we take an example from ref.\citep{bonaccorso2013sorting} where Au nanoparticles (AuNPs) are fractionated by centrifugation. We consider the AuNP sample has a log-normal radius distribution with mean value 7.9 nm and standard deviation 5.1 nm (see Table 1 and Figure 2a in ref.\citep{bonaccorso2013sorting}, note that in ref.\citep{bonaccorso2013sorting} size was characterized using diameter whereas it's radius here). The frequency distribution of the radius is shown in Fig.\ref{fig15} with a cut at 25 nm. The desired radius range is 5-15 nm. As shown in Sec.\ref{sec3.3.2}, a two-step protocol can be adopted. In the first step, the centrifugation parameters (time or rpm) should be calculated using the largest desired size (i.e. 15 nm). In the second step, the unoptimized way is to calculate the centrifugation parameters using the smallest desired size (i.e. 5 nm). However, as shown in Sec.\ref{sec3.3.2}, adding an offset value ($\Delta a$) to the smallest desired size to calculate the centrifugation parameters can lead to lower impurity percentage (optimized way). Note that the possible value of $\Delta a$ in this case is between 0 to 10 nm. By numerically solving the set of linear convective equations Eq.\ref{eq4}, the effect of adding an offset value $\Delta a$ on impurity percentage and yield is shown in Fig.\ref{fig16}. Note that Fig.\ref{fig16} is constructed assuming an homogeneous solution before each step of the procedure (i.e. homogeneous centrifugation), as this is more easy and practical for multicascade centrifugation. From this reference map, we can see that by choosing $\Delta a$ as 3 nm the impurity percentage drops from 29\% to 17\% whereas the yield (defined in Sec.\ref{sec3.3.2}) drops from 65\% to 50\%. The size frequency distributions of the fractionated sample by unoptimized or optimized (choosing $\Delta a$ as 3 nm) two-step procedures are shown in Fig.\ref{fig17}. From Fig.\ref{fig17}, it is seen that the distribution from the optimized procedure has a larger fraction in the desired range than that from the unoptimized procedure, which means the fractionated sample from the optimized procedure has lower impurity percentage. The mean values of the distributions shown in Fig.\ref{fig17} are 7.40 nm for optimized procedure and 6.57 nm for unoptimized procedure. The standard deviations are 2.46 nm for optimized procedure and 2.43 nm for unoptimized procedure. To calculate the centrifugation parameters of the two-step procedure, a calculator app is implemented in Matlab. By choosing the medium liquid as PEG-200 and some operating conditions from ref.\citep{bonaccorso2013sorting}, the rpm needed is about 12Krpm in step 1 and 22Krpm in step 2 if it is chosen to centrifuge for 2 hours in both steps, as shown in Fig.\ref{fig18}.    


\subsection{Fractionation of graphene (anisotropic particles)} \label{sec:graphene}
Liquid-phase exfoliation (LPE) is a promising method \citep{coleman2011two,yi2015review} to produce 2D nanomaterials, such as  graphene \citep{hernandez2008high,khan2010high,li2020mechanisms}, MoS\textsubscript{2} \citep{o2012preparation,bang2014effective} or BN \citep{mukhopadhyay2017deciphering,wang2020insight}. The flakes produced with LPE tend to be polydispersed in lateral size and thickness \citep{ogilvie2019size,chacham2020controlling}, with both variables following approximately log-normal distributions \citep{chacham2020controlling,kouroupis2014fragmentation}. Fractionation of 2D nanomaterials by centrifugation is very common \citep{sun2010monodisperse,khan2012size,bonaccorso2013sorting,backes2014edge,backes2016production}.
  
  
In the following we refer to graphene, but the results are equally applicable to other 2D nanomaterials. Modelling a graphene flake as a disk-like particle of diameter $d$ and thickness $L$, the settling velocity is (see Appendix B for derivation)
\begin{equation}
    \label{eq9}
    q=\frac{\pi}{27.48\mu}(\rho_p - \rho_l) g_e dL.
\end{equation}



The settling velocity $q$ of a disk is proportional to the product of lateral size $d$ and thickness $L$. If both $d$ and $L$ are log-normally distributed, $q$ also follows log-normal distribution assuming $d$ and $L$ are independent. However, statistical analysis of large amount nanoflakes produced by LPE shows that the average values of lateral size and thickness are correlated by a power low $<d> \sim <L>^r$, with the exponent $r$ depending on the type of the material and the exfoliation procedure \citep{chacham2020controlling,ogilvie2019size}. Based on this relationship, \citet{ogilvie2019size} proposed the following expression for the average settling velocity of 2D nanosheets:
\begin{equation}
    \label{eq12}
    \langle q \rangle=\frac{(\rho_p - \rho_l)k^2 \langle \mathcal{L} \rangle^m g_e}{6 \pi \mu}.
\end{equation}
In Eq.\ref{eq12}, $\langle \mathcal{L} \rangle$ is the average layer number of the nanosheets, and $k$ is a shape factor related to the aspect ratios of the nanosheets with the dimension of length. An average and idealised value for the exponent $m$ is 2.5, covering a broad initial distributions of $\langle \mathcal{L} \rangle$ of different 2D materials. The values of $k$ and $m$ can be determined by a calibration experiment \citep{ogilvie2019size}. Then, Eq.\ref{eq12} can be utilized to calculate the parameters needed for centrifugation (i.e. equivalent $g$-force and centrifugation time).

We will now explain how to use the one-dimensional model (Eq.\ref{eq4}) to predict the outcome of the fractionation of nanosheets by centrifugation. Suppose nanosheets with layer number between $\mathcal{L}_1$ and $\mathcal{L}_2$ ($\mathcal{L}_1 < \mathcal{L}_2$) are desired. First, a calibration experiment can be done to acquire the values of $k$ and $m$ in Eq.\ref{eq12}. The frequency distribution of layer number of the nanosheets ($\mathcal{L}$) can be obtained by atomic force microscopy (AFM). Second, the frequency distribution of the settling velocity can be obtained once the physical properties of the nanosheets (i.e. density) and solvent (i.e. density and viscosity) and the equivalent $g$-force used in the centrifugation are known, according to Eq.\ref{eq12}. The two settling velocities $q_1$ and $q_2$ of nanosheets with layer number $\mathcal{L}_1$ and $\mathcal{L}_2$, respectively, can be calculated. Third, the centrifugation times used in the two steps can be chosen as $t_1=H/q_2$ and $t_2=H/q_1$, respectively, and $H$ is the filling height of the dispersion in the vial. After centrifuging for the duration of $t_1$, the sediment can be discarded and the supernatant can be centrifuged again for $t_2$, and the sediment from the second step can be collected as the fractionated sample. By solving the one-dimensional model (Eq.\ref{eq4}) with corresponding parameters (i.e. the centrifugation times) and initial conditions (i.e. band or homgeneous sedimentation, initial frequency distribution), the frequency distribution of the fractionated sample can be obtained. Based on this distribution, the full statistics of the fractionated sample (e.g. range, mean value and standard deviation of the layer number) can be predicted. Finally, to increase the purity of the fractionated sample, the centrifugation time in the second step can be lowered to $t_2=H/(q_1+\Delta q)$. By solving the one-dimensional model, a reference map like Fig.\ref{fig16} of the impurity percentage and yield when choosing different $\Delta q$ can be constructed. Then, the appropriate value of $\Delta q$ to be used in the centrifugation process can be chosen based on the desired purity and yield.

Similarly, the model and corresponding experimental protocol can be used to fractionate nanosheets by lateral size. Suppose a sample of nanosheets with approximately the same thickness whereas the lateral size is log-normally distributed with mean as \SI{0.1}{\micro\meter} and standard deviation as \SI{0.1}{\micro\meter}, to be fractionated by lateral size, and the desired range of lateral size is from \SI{0.025}{\micro\meter} to \SI{0.045}{\micro\meter}. Since the nanosheets have the same thickness, $q \sim d$ according to Eq.\ref{eq9}. Therefore $q$ also follows log-normal distribution. Take the settling velocity of a nanosheet whose lateral size is \SI{1}{\micro\meter} as characteristic velocity to non-dimensionalize $q$, the system is converted to the one studied in Sec.\ref{results}. If the two-step approach shown in Sec.\ref{sec3.3.2} is adopted to fractionate the sample, and the centrifugation time in the second step ($t_2$) is based on the settling velocity of the nanosheet whose lateral size is \SI{0.025}{\micro\meter}, the impurity percentage of the final product would be about 5\% using band sedimentation (assuming initially the height of particle-laden layer is 10\% of the vial height), and would be about 50\% using homogeneous sedimentation, according to Fig.\ref{fig12}(b). To lower the impurity percentage, $t_2$ could be lowered. For example, the impurity percentage would be 0 ideally in band sedimentation if $t_2$ is based on the settling velocity of the nanosheet whose lateral size is \SI{0.02778}{\micro\meter}, and would be about 45\% in homogeneous sedimentation if $t_2$ is based on the settling velocity of the nanosheet whose lateral size is \SI{0.035}{\micro\meter}, according to Fig.\ref{fig13}. In the meantime, the yield would decrease from 86\% to 80\% in band sedimentation, and would decrease from 21.5\% to 19\% in homogeneous sedimentation.      
\section{Conclusions}\label{sec5}
In this paper, a mathematical model is used to study the multicascade (multistep) sedimentation or centrifugation of polydisperse particle systems with an initial log-normal size distribution, under the main assumption that the suspension is dilute.  Two cases are considered: band sedimentation, where the particles are initially deposited in a small slab at the top of the clear fluid, and homogeneous sedimentation, where the particles are evenly dispersed in the vial initially. The model enables to predict the time evolution of the size frequency distributions in the supernatant and in the sediment layer, and the conditions for optimal sorting of an initial size distribution in distinct particle size classes. Fractionation of metallic nanoparticles and fractionation of graphene nanosheets are taken as examples to illustrate the practical application of the theory. The main conclusions of our analysis  are as follows.

In band sedimentation, the frequency distribution curves in the supernatant and in the sediment layer overlap only slightly at each time  (see Fig.\ref{fig7}(a)), meaning there is a clear distinction between the size classes in the supernatant and in the sediment layer. Therefore, in the absence of convective or diffusive mixing, band sedimentation is preferable over homogeneous sedimentation to fractionate particles by size. In applications where the purity of the sample is paramount, therefore, band sedimentation could offer significant advantages. The main drawback of band sedimentation is the lower yield and the fact that, in practice, it is more difficult to controllably place a layer of particle-rich fluid over the clear fluid without incurring in gravitational instabilities that lead to fluid mixing. Methods to overcome this practical issue are using solvent with density gradient \citep{velegol2009rayleigh} and adding a buffer layer liquid between the suspension and clear fluid \citep{coll1986use}. Given the significant advantages of band sedimentation in terms of purity, new methods to prevent mixing when using band sedimentation should  be investigated in the future. 
    
We have proposed and analyzed a rational protocol to isolate a certain particle size range. The protocol, described in Sec. \ref{sec3.3}, involves two characteristic centrifugation or sedimentation times: $t_{min}=H/q_{max}$ and $t_{max}=H/q_{min}$, where $H$ is the height of the free surface of the liquid with respect to the bottom of the vial,  $q_{max}$ is the velocity of the largest (fastest) particles and $q_{min}$ is the velocity of the smallest (slowest) particles. The protocol involves sedimenting or centrifuging for time $t_{min}=H/q_{max}$, transferring the supernatant to a new vial and centrifuging for a longer time. Then the process is repeated a number of steps. In analyzing this method we have found that, surprisingly, a two-step method is not worse than a multi-step method when the objective is to isolate a given particle size range  (see Fig.\ref{fig12}). Therefore, to isolate a size range that lies in the middle of the size distribution, only 2 steps could be used. Using this result, it can be easily calculated that to fractionate an entire initial size distribution into $N_c$ classes, only $N_c-1$ steps could be used, starting from the largest particle class and then proceeding towards the smallest particles, `slicing' progressively the size distribution. Given that such steps could be automatized, this offers the opportunity for fast sorting of particulate suspensions based on a limited number of steps. Furthermore, we have found that shifting the time by a small amount can lead to an even more precise selection of the size class, i.e. the attainment of a sample with reduced amount of ``impurities'' (particles that are selected but are not in the desired particle class). In both the initial protocol and in the improved one there is, in general, a trade-off between yield and amount of impurities.  
    
The advantage of the methods we propose is that they are based on the equations of motion of the particles in the fluid, and therefore using the parameters of the paper will lead to exact predictions in the very dilute limit in which the sedimentation velocity of one particle does not depend on the presence of the other particles. The analysis can be used as a theoretical guideline to design or modify sedimentation and centrifugation protocols for more realistic situations where the suspension is not truly dilute. For instance, we envision application in the fractionation of graphene particles, and have provided practical guidelines for this case (see Sec. \ref{sec:graphene}).

The digitalisation of laboratory procedures means that exact algorithms will be required to replace choices based on empiricism, so the availability of exact formulas and accurate quantitative predictions for centrifugation/sedimentation will become increasingly  useful, both in analytical laboratories and in plant operations. In the future, we will investigate with the help of high-resolution, particle-resolved simulations how well the dilute theory is able to describe the evolution of the size distribution when the suspension is not so dilute that hydrodynamic interactions can be neglected and the flow microphysics leading to convective mixing in band sedimentation/centrifugation. 

\section*{Acknowledgements}
This research has received funding from the European Research Council (ERC) under the European Union's Horizon 2020 research and innovation programme (grant agreement No. 715475).

\section*{Appendix}
\section*{Appendix A: Derivation of Eq.\ref{eq11}}\label{AB}
Since Eq.\ref{eq2} is linear, it can be solved analytically, and the solution under initial condition Eq.\ref{eq3} is:
\begin{align*}
    N(x,t,q)=N_0(x-qt,q).
\end{align*}
Based on this analytical solution, the number of particles with settling velocity $q$ in the supernatant at time $t$ is:
\begin{align*}
    n^s_q(t) & = \int_0^H N(x,t;q)dx = \int_0^H f_0(q)n_0(x-qt)dx \\
    & =f_0(q)\int_{-qt}^{H-qt}n_0(x-qt)d(x-qt) = f_0(q)\int_{-qt}^{H-qt}n_0(u)du.
\end{align*}

For band sedimentation, based on the expression for initial total particle number density $n_0(x)$, we have:
\begin{align*}
    n_{q}^{s}(t)=\begin{cases}
    f_0(q)\times 1\times h=hf_0(q) & \text{$H-qt\geq h$} \\
    f_0(q)\times 1 \times(H-qt)=(H-qt)f_0(q) & \text{$0 < H-qt < h$} \\
    0 & \text{$H-qt \leq 0$}
    \end{cases}
    .
\end{align*}
Here $h$ is the thickness of the initial particle-laden layer, $H$ is the filling height of the dispersion, and the value of $n_0$ is chosen as $1$.

For homogeneous sedimentation, we have:
\begin{align*}
    n_{q}^{s}(t)=\begin{cases}
    (H-qt)f_0(q) & \text{$H-qt>0$} \\
    0 & \text{$H-qt \leq 0$}
    \end{cases}
    .
\end{align*}

Based on the expression for $n_q^s(t)$, we know that at time $t$ the number of particles of $q$ larger than $H/t$ is $0$ in the supernatant. Since the largest value of $q$ in the sample is $Q$, the largest value of $q$ in the supernatant at time $t$ is:
\begin{align*}
    q_{max}^s = \begin{cases}
    H/t & \text{$t > H/Q$} \\
    Q & \text{$t \leq H/Q$}
    \end{cases}
    .
\end{align*}

\section*{Appendix B: Derivation of Eq.\ref{eq9}}\label{AA}
Consider a thin disk with lateral size $d$ and thickness $L$ settling in a viscous liquid under an external force field with equivalent $g$-force $g_e$. The body force on this disk is:
\begin{align*}
    F_b = (\rho_p - \rho_l) \frac{\pi}{4} d^2 L g_e.
\end{align*}
Here, $\rho_p$ and $\rho_l$ are densities of the particle and liquid, respectively. The drag force on the disk can be expressed as:
\begin{align*}
    F_d = f \cdot 3\pi\mu_l d_e q,
\end{align*}
where $\mu_l$ is the dynamic viscosity of the liquid, $d_e$ is the equivalent-volume diameter of the particle which equals $(3Ld^2/2)^{1/3}$ for a disk particle, $q$ is the settling velocity, and $f$ is the correction factor due to the non-spherical shape \citep{loth2008drag}. For a thin disk with aspect ratio $E=L/d$, the correction factor $f$ is $\frac{8E^{-1/3}}{3\pi}$ when the disk settles in the direction parallel to its axis of symmetry (broadwise), and is $\frac{16E^{-1/3}}{9\pi}$ when the disk settles in the direction perpendicular to its axis (edgewise) \citep{loth2008drag}. To account for the rotational Brownian motion, the correction factor is averaged over all orientations, which is $\frac{2E^{-1/3}}{\pi}$. This gives the drag force as:
\begin{align*}
    F_d = 6.87\mu_l qd.
\end{align*}
By equating the body force and drag force on the disk, the settling velocity of a thin disk particle is
\begin{align*}
    q=\frac{\pi}{27.48\mu_l}(\rho_p-\rho_l)g_edL.
\end{align*}

\bibliographystyle{elsarticle-num-names} 
\bibliography{references}

\end{document}